\documentclass[10pt,a4paper,onecolumn]{article}
\usepackage{marginnote}
\usepackage{graphicx}
\usepackage{xcolor}
\usepackage{authblk,etoolbox}
\usepackage{titlesec}
\usepackage{calc}
\usepackage{tikz}
\usepackage{hyperref}
\hypersetup{colorlinks,breaklinks,
            urlcolor=[rgb]{0.0, 0.5, 1.0},
            linkcolor=[rgb]{0.0, 0.5, 1.0}}
\usepackage{caption}
\usepackage{tcolorbox}
\usepackage{amssymb,amsmath}
\usepackage{ifxetex,ifluatex}
\usepackage{seqsplit}
\usepackage{xstring}

\usepackage{fixltx2e} 
\usepackage[
  backend=biber,
]{biblatex}
\bibliography{paper.bib}

\let\textttOrig=\texttt
\def\texttt#1{\expandafter\textttOrig{\seqsplit{#1}}}
\renewcommand{\seqinsert}{\ifmmode
  \allowbreak
  \else\penalty6000\hspace{0pt plus 0.02em}\fi}

\makeatletter
\let\href@Orig=\href
\def\href@Urllike#1#2{\href@Orig{#1}{\begingroup
    \def\Url@String{#2}\Url@FormatString
    \endgroup}}
\def\href@Notdoi#1#2{\def\tempa{#1}\def\tempb{#2}%
  \ifx\tempa\tempb\relax\href@Urllike{#1}{#2}\else
  \href@Orig{#1}{#2}\fi}
\def\href#1#2{%
  \IfBeginWith{#1}{https://doi.org}%
  {\href@Urllike{#1}{#2}}{\href@Notdoi{#1}{#2}}}
\makeatother

\usepackage[top=3.5cm, bottom=3cm, right=1.5cm, left=1.0cm,
            headheight=2.2cm, reversemp, includemp, marginparwidth=4.5cm]{geometry}

\titleformat{\section}
  {\normalfont\sffamily\Large\bfseries}
  {}{0pt}{}
\titleformat{\subsection}
  {\normalfont\sffamily\large\bfseries}
  {}{0pt}{}
\titleformat{\subsubsection}
  {\normalfont\sffamily\bfseries}
  {}{0pt}{}
\titleformat*{\paragraph}
  {\sffamily\normalsize}

\usepackage{fancyhdr}
\makeatletter
\let\ps@plain\ps@fancy
\fancyheadoffset[L]{4.5cm}
\fancyfootoffset[L]{4.5cm}

\definecolor{linky}{rgb}{0.0, 0.5, 1.0}

\newtcolorbox{repobox}
   {colback=red, colframe=red!75!black,
     boxrule=0.5pt, arc=2pt, left=6pt, right=6pt, top=3pt, bottom=3pt}

\newcommand{\ExternalLink}{%
   \tikz[x=1.2ex, y=1.2ex, baseline=-0.05ex]{%
       \begin{scope}[x=1ex, y=1ex]
           \clip (-0.1,-0.1)
               --++ (-0, 1.2)
               --++ (0.6, 0)
               --++ (0, -0.6)
               --++ (0.6, 0)
               --++ (0, -1);
           \path[draw,
               line width = 0.5,
               rounded corners=0.5]
               (0,0) rectangle (1,1);
       \end{scope}
       \path[draw, line width = 0.5] (0.5, 0.5)
           -- (1, 1);
       \path[draw, line width = 0.5] (0.6, 1)
           -- (1, 1) -- (1, 0.6);
       }
   }

\patchcmd{\@maketitle}{center}{flushleft}{}{}
\patchcmd{\@maketitle}{center}{flushleft}{}{}
\patchcmd{\@maketitle}{\LARGE}{\LARGE\sffamily}{}{}
\def\maketitle{{%
  
  \AB@maketitle}}
\makeatletter
\renewcommand\AB@affilsepx{ \protect\Affilfont}
\renewcommand\AB@affilnote[1]{{\bfseries #1}\hspace{3pt}}
\renewcommand{\affil}[2][]%
   {\newaffiltrue\let\AB@blk@and\AB@pand
      \if\relax#1\relax\def\AB@note{\AB@thenote}\else\def\AB@note{#1}%
        \setcounter{Maxaffil}{0}\fi
        \begingroup
        \let\href=\href@Orig
        \let\texttt=\textttOrig
        \let\protect\@unexpandable@protect
        \def\thanks{\protect\thanks}\def\footnote{\protect\footnote}%
        \@temptokena=\expandafter{\AB@authors}%
        {\def\\{\protect\\\protect\Affilfont}\xdef\AB@temp{#2}}%
         \xdef\AB@authors{\the\@temptokena\AB@las\AB@au@str
         \protect\\[\affilsep]\protect\Affilfont\AB@temp}%
         \gdef\AB@las{}\gdef\AB@au@str{}%
        {\def\\{, \ignorespaces}\xdef\AB@temp{#2}}%
        \@temptokena=\expandafter{\AB@affillist}%
        \xdef\AB@affillist{\the\@temptokena \AB@affilsep
          \AB@affilnote{\AB@note}\protect\Affilfont\AB@temp}%
      \endgroup
       \let\AB@affilsep\AB@affilsepx
}
\makeatother

\renewcommand\Affilfont{\sffamily\small\mdseries}
\ifnum 0\ifxetex 1\fi\ifluatex 1\fi=0 
  \usepackage[T1]{fontenc}
  \usepackage[utf8]{inputenc}
  \usepackage{lmodern}

\else 
  \ifxetex
    \usepackage{mathspec}
  \else
    \usepackage{fontspec}
  \fi
  \defaultfontfeatures{Ligatures=TeX,Scale=MatchLowercase}

\fi
\IfFileExists{upquote.sty}{\usepackage{upquote}}{}
\IfFileExists{microtype.sty}{%
\usepackage{microtype}
\UseMicrotypeSet[protrusion]{basicmath} 
}{}

\usepackage{hyperref}
\hypersetup{unicode=true,
            pdftitle={pyro: a framework for hydrodynamics explorations and prototyping},
            pdfborder={0 0 0},
            breaklinks=true}
\let\addcontentslineOrig=\addcontentsline
\def\addcontentsline#1#2#3{\bgroup
  \let\texttt=\textttOrig\addcontentslineOrig{#1}{#2}{#3}\egroup}
\let\markbothOrig\markboth
\def\markboth#1#2{\bgroup
  \let\texttt=\textttOrig\markbothOrig{#1}{#2}\egroup}
\let\markrightOrig\markright
\def\markright#1{\bgroup
  \let\texttt=\textttOrig\markrightOrig{#1}\egroup}

\usepackage{graphicx,grffile}
\makeatletter
\def\maxwidth{\ifdim\Gin@nat@width>\linewidth\linewidth\else\Gin@nat@width\fi}
\def\maxheight{\ifdim\Gin@nat@height>\textheight\textheight\else\Gin@nat@height\fi}
\makeatother
\setkeys{Gin}{width=\maxwidth,height=\maxheight,keepaspectratio}
\IfFileExists{parskip.sty}{%
\usepackage{parskip}
}{
\setlength{\parindent}{0pt}
\setlength{\parskip}{6pt plus 2pt minus 1pt}
}
\ifx\paragraph\undefined\else
\let\oldparagraph\paragraph
\renewcommand{\paragraph}[1]{\oldparagraph{#1}\mbox{}}
\fi
\ifx\subparagraph\undefined\else
\let\oldsubparagraph\subparagraph
\renewcommand{\subparagraph}[1]{\oldsubparagraph{#1}\mbox{}}
\fi

\title{pyro: a framework for hydrodynamics explorations and prototyping}

        \author[1]{Alice Harpole}
          \author[1]{Michael Zingale}
          \author[2]{Ian Hawke}
          \author[3]{Taher Chegini}

      \affil[1]{Department of Physics and Astronomy, Stony Brook University}
      \affil[2]{University of Southampton}
      \affil[3]{University of Houston}
  \date{\vspace{-5ex}}

\begin{document}
\maketitle

\marginpar{
  \sffamily\small

  {\bfseries DOI:} \href{https://doi.org/10.21105/joss.01265}{\color{linky}{10.21105/joss.01265}}

  \vspace{2mm}

  {\bfseries Software}
  \begin{itemize}
    \setlength\itemsep{0em}
    \item \href{https://github.com/openjournals/joss-reviews/issues/588}{\color{linky}{Review}} \ExternalLink
    \item \href{https://github.com/python-hydro/pyro2}{\color{linky}{Repository}} \ExternalLink
    \item \href{http://dx.doi.org/10.5281/zenodo.2575565}{\color{linky}{Archive}} \ExternalLink
  \end{itemize}

  \vspace{2mm}

  {\bfseries Submitted:} 15 February 2019\\
  {\bfseries Published:} 22 March 2019

  \vspace{2mm}
  {\bfseries License}\\
  Authors of papers retain copyright and release the work under a Creative Commons Attribution 4.0 International License (\href{http://creativecommons.org/licenses/by/4.0/}{\color{linky}{CC-BY}}).
}

\section{Summary}\label{summary}

\texttt{pyro} is a Python-based simulation framework designed for ease
of implementation and exploration of hydrodynamics methods. It is built
in a object-oriented fashion, allowing for the reuse of the core
components and fast prototyping of new methods.

The original goal of \texttt{pyro} was to learn hydrodynamics methods
through example, and it still serves this goal. At Stony Brook,
\texttt{pyro} is used with new undergraduate researchers in our group to
introduce them to the ideas of computational hydrodynamics. But the
current framework has evolved to the point where \texttt{pyro} is used
for prototyping hydrodynamics solvers before implementing them into
science codes. An example of this is the 4th-order compressible solver
built on the ideas of spectral deferred corrections (the
\texttt{compressible\_sdc} solver). This implementation was used as the
model for the development of higher-order schemes in the Castro
hydrodynamics code (Almgren et al. 2010). The low Mach number
atmospheric solver (\texttt{lm\_atm}) is based on the Maestro code
(Nonaka et al. 2010) and the \texttt{pyro} implementation will be used
to prototype new low Mach number algorithms before porting them to
science codes.

In the time since the first \texttt{pyro} paper (Zingale 2014), the code
has undergone considerable development, gained a large number of
solvers, adopted unit testing through pytest and documentation through
sphinx, and a number of new contributors. \texttt{pyro}'s functionality
can now be accessed directly through a \texttt{Pyro()} class, in
addition to the original commandline script interface. This new
interface in particular allows for easy use within Jupyter notebooks. We
also now use HDF5 for output instead of Python's \texttt{pickle()}
function. Previously, we used Fortran to speed up some
performance-critical portions of the code. These routines could be
called by the main Python code by first compiling them using
\texttt{f2py}. In the new version, we have replaced these Fortran
routines by Python functions that are compiled at runtime by
\texttt{numba}. Consequently, \texttt{pyro} is now written entirely in
Python.

The current \texttt{pyro} solvers are:

\begin{itemize}
\item
  linear advection (including a second-order unsplit CTU scheme, a
  method-of-lines piecewise linear solver\(^\star\), a 4th-order
  finite-volume scheme\(^\star\), a WENO method\(^\star\), and advection
  with a non-uniform velocity field\(^\star\))
\item
  compressible hydrodynamics (including a second-order unsplit CTU
  scheme, a method-of-lines piecewise linear solver\(^\star\), and two
  4th-order finite-volume schemes, one with Runge-Kutta integration and
  the other using a spectral deferred corrections method\(^\star\))
\item
  diffusion using a second-order implicit discretization
\item
  incompressible hydrodynamics using a second-order approximate
  projection method.
\item
  low Mach number atmospheric solver\(^\star\), using an approximate
  projection method.
\item
  shallow water equations solver\(^\star\)
\end{itemize}

(solvers since the first \texttt{pyro} paper are marked with a
\(^\star\)). Also, new is support for Lagrangian tracer particles, which
can be added to any solver that has a velocity field.

\section{Acknowledgements}\label{acknowledgements}

The work at Stony Brook was supported by DOE/Office of Nuclear Physics
grant DE-FG02-87ER40317 and DOE grant DE-SC0017955.

\section*{References}\label{references}
\addcontentsline{toc}{section}{References}

\hypertarget{refs}{}
\hypertarget{ref-castro}{}
Almgren, A. S., V. E. Beckner, J. B. Bell, M. S. Day, L. H. Howell, C.
C. Joggerst, M. J. Lijewski, A. Nonaka, M. Singer, and M. Zingale. 2010.
``CASTRO: A New Compressible Astrophysical Solver. I. Hydrodynamics and
Self-gravity.'' \emph{Astrophysical Journal} 715 (June): 1221--38.
doi:\href{https://doi.org/10.1088/0004-637X/715/2/1221}{10.1088/0004-637X/715/2/1221}.

\hypertarget{ref-maestro}{}
Nonaka, A., A. S. Almgren, J. B. Bell, M. J. Lijewski, C. M. Malone, and
M. Zingale. 2010. ``MAESTRO: An Adaptive Low Mach Number Hydrodynamics
Algorithm for Stellar Flows.'' \emph{Astrophysical Journal Supplement}
188 (June): 358--83.
doi:\href{https://doi.org/10.1088/0067-0049/188/2/358}{10.1088/0067-0049/188/2/358}.

\hypertarget{ref-pyroI}{}
Zingale, M. 2014. ``pyro: A teaching code for computational
astrophysical hydrodynamics.'' \emph{Astronomy and Computing} 6
(October): 52--62.
doi:\href{https://doi.org/10.1016/j.ascom.2014.07.003}{10.1016/j.ascom.2014.07.003}.

\end{document}